\documentclass[conference]{IEEEtran}
\usepackage{graphicx}
\usepackage{latexsym}
\usepackage{amssymb}
\usepackage{amsmath}
\usepackage{cite}
\usepackage{subfigure}
\usepackage{stfloats}

\newtheorem{lemma}{Lemma}

\DeclareMathOperator{\E}{\mathbb{E}}

\newcommand{\B}[1]{{\pmb{#1}}}
\newcommand{\EX}[1]{\E\left\{{#1}\right\}}

\newcommand{\MGF}[2]{\Phi_{#1}\left({#2}\right)}
\newcommand{\EXs}[2]{\E_{{#1}}\left\{{#2}\right\}}
\newcommand{\PDF}[2]{p_{{#1}}\left({#2}\right)}
\newcommand{\CDF}[2]{F_{{#1}}\left({#2}\right)}

\newcommand{\Rsum}{C_{\text{sum}}}
\newcommand{\FN}[1]{\left\|{#1}\right\|_\mathrm{F}}
\newcommand{\CG}[2]{\mathcal{CN}\left({#1},{#2}\right)}
\newcommand{\CGV}[3]{\tilde{\mathcal{N}}_{#1}\left({#2},{#3}\right)}

\newcommand{\Bessel}[2]{\mathcal{K}_{{#1}}\left({#2}\right)}

\newcommand{\Source}[1]{\mathsf{S}_{#1}}
\newcommand{\Relay}{\mathsf{R}}
\newcommand{\Ps}{\mathcal{P}_s}
\newcommand{\y}[1]{\B{y}_{\mathsf{{#1}}}}
\newcommand{\h}[1]{h_{#1}}
\newcommand{\hnew}{\B{h}}
\newcommand{\s}[1]{\B{s}_{#1}}
\newcommand{\x}[1]{\B{x}_{#1}}
\newcommand{\nR}{\B{n}_{\mathsf{R}}}
\newcommand{\n}[1]{\B{n}_{#1}}
\newcommand{\z}[1]{\B{z}_{#1}}
\newcommand{\CP}[1]{\Omega_{#1}}
\newcommand{\snri}[1]{\gamma_{#1}}
\newcommand{\snravei}[1]{\bar\gamma_{#1}}
\newcommand{\Gain}{G}
\newcommand{\snrave}{\bar\gamma}
\newcommand{\R}[1]{\B{R}_{#1}}

\title{Average Sum-Rate of Distributed Alamouti Space--Time Scheme in Two-Way Amplify-and-Forward Relay Networks}
\author{\authorblockN{Trung Q. Duong$^{\dag}$, Chau Yuen$^\ddag$, Hans-J{\"{u}}rgen Zepernick$^{\dag}$, and Xianfu Lei$^\S$}
\authorblockA{\small $^{\dag}$Blekinge Institute of
Technology, SE-371 79, Karlskrona, Sweden \\ $^\ddag$Singapore
University of Technology and Design, Singapore \\ $^\S$Southwest
Jiaotong University, Chengdu 610031, China \\ E-mail: dqt@bth.se,
yuenchau@sutd.edu.sg, hjz@bth.se, xflei81@yahoo.com.cn}}

\begin{document}
\maketitle

%
\begin{abstract}
In this paper, we propose a distributed Alamouti space-time code
(DASTC) for two-way relay networks employing a \emph{single}
amplify-and-forward (AF) relay. We first derive closed-form
expressions for the approximated average sum-rate of the proposed
DASTC scheme. Our analysis is validated by a comparison against the
results of Monte-Carlo simulations. Numerical results verify the
effectiveness of our proposed scheme over the conventional DASTC
with one-way communication.
\end{abstract}


\section{Introduction}

Relay networking has been considered as an efficient approach to
increase the communication range of wireless systems. However,
dual-hop half-duplex relay networks lose half of the throughput
compared to the direct communication due to the fact that the relay
cannot transmit and receive simultaneously. To overcome this
drawback, a two-way (or bi-directional) relay network has been
presented in \cite{RW:07:JSAC}, where two nodes, namely $\Source{1}$
and $\Source{2}$, transmit simultaneously to the relay $\Relay$ in
the first hop, and in the second hop the relay $\Relay$ forwards its
received signals to both terminals $\Source{1}$ and $\Source{2}$.
With this strategy, this loss can be remarkably compensated.

As a result, two-way relay networks have gained great attention in
the research community (e.g., see
\cite{RW:07:JSAC,PY:07:CL,UK:08:EURASIP,DHB:09:IEICE,LLV:10:WCOM}).
In view of the related work of two-way relay networks, the upper and
lower bounds for average sum-rate have been investigated in
\cite{RW:07:JSAC}. The performance of two-way amplify-and-forward
(AF) relay with network coding has been investigated in
\cite{PY:07:CL}. By giving up the strict separation of downlink and
uplink signals through either time or frequency division duplex, a
two-hop relaying, namely space division duplex relaying, is proposed
in \cite{UK:08:EURASIP}. The exact closed-form expressions of error
probability, average sum-rate for two-way AF relays have been
presented in \cite{DHB:09:IEICE,LLV:10:WCOM}.

Distributed Alamouti space-time code (DASTC) with AF relays
originally applied for one-way relay networks \cite{Duong:08:VTC}
has recently extended to two-way systems (see, e.g.,
\cite{HTHC:08:VTCSpring,HTHC:09:WCOM,ECYL:09:VTCFall}). In
particular, a two-way relaying scheme where two sources equipped
with two antennas transmit Alamouti code through the help of an AF
relay has been proposed in \cite{HTHC:08:VTCSpring,HTHC:09:WCOM}.
Upper and lower bounds of average sum-rate have also been derived
for this particular bi-directional relaying system. Another system
in which the burden of deploying multiple antennas on sources is
transferred to the relay has been proposed in
\cite{ECYL:09:VTCFall}. By assuming that the relay is equipped with
two antennas and each source is equipped with a single antenna, an
upper bound of symbol error probability has been obtained. As can be
observed from all the above schemes, either relay or source requires
multiple antennas which may be prohibited in practical
implementations due to the high demand for low-cost and small-size
portable devices.

To alleviate this requirement and make the system realistic, a DASTC
scheme which all terminals having only one antenna has been
presented for one-way AF relay networks
\cite{NBK:04:JSAC,DNHN:09:WPC}. However, this system still faces
loss in spectral efficiency. Hence, in this paper, we propose a
DASTC scheme for two-way AF relay networks which significantly
recovers this loss. More importantly, unlike the analysis work in
\cite{HTHC:08:VTCSpring,HTHC:09:WCOM} where the bounds of average
sum-rate have been shown, we have derived an asymptotically tight
approximation for the average sum-rate of the proposed two-way
scheme. The final expression is given in the form of Fox's
H-function which enables us to investigate the performance of the
proposed scheme. In addition, we also provide the numerical results
to verify the correctness of our analysis.

The rest of this paper is organized as follows. In
Section~\ref{sec:SystemModel}, we introduce the system model of the
proposed DASTC for two-way AF relay networks. Then, in
Section~\ref{sec:SumRate}, we derive the tight approximation for the
average sum-rate of the proposed scheme. Numerical results are shown
in Section~\ref{sec:Result} to validate the analysis. Finally, we
conclude the paper in Section~\ref{sec:Conclusions}.

\emph{Notation:} Throughout the paper, we shall use the following
notation. Vector is written as bold lower case letter and matrix is
written as bold upper case letter. The superscripts $\ast$ and
$\dag$ stand for the complex conjugate and transpose conjugate,
respectively. $\B{I}_n$ represents the $n \times n$ identity matrix.
$\FN{\B{A}}$ denotes Frobenius norm of the matrix $\B{A}$ and
$\left|x\right|$ indicates the envelope of $x$. $\EXs{x}{.}$ is the
expectation operator over the random variable $x$. A complex
Gaussian distribution with mean $\mu$ and variance $\sigma^2$ is
denoted by $\mathcal{CN}(\mu,\sigma^2)$. Let us denote
$\CGV{m}{\B{m}}{\B{\Sigma}}$ as a complex Gaussian random vector
with mean vector $\B{m}$ and covariance matrix $\B{\Sigma}$.
$\Gamma\left( {a,x} \right)$ is the incomplete gamma function
defined as $ \Gamma \left( {a,x} \right) = \int_x^\infty {t^{a - 1}
e^{ - t} dt}$ and $\Bessel{n}{.}$ is the $n$th-order modified Bessel
function of the second kind.

\section{System Model of Distributed Alamouti Space-Time Codes with Two-Way AF Relaying} \label{sec:SystemModel}

\begin{figure}[t]
    \centerline{\includegraphics[width=0.35\textwidth]{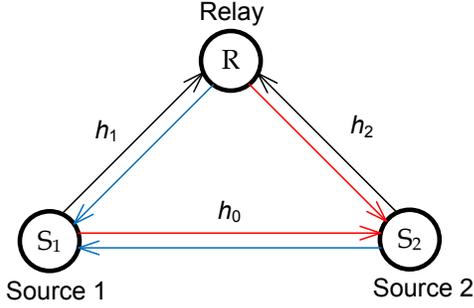}}
    \caption{
        System model.
    }
    \label{fig:sys}
\end{figure}

In this section, we introduce the system model of the proposed DASTC
scheme with two-way AF relaying. Consider a wireless AF relay
network consisting of three terminals as shown in
Fig.~\ref{fig:sys}. Two sources, namely $\Source{1}$ and
$\Source{2}$, exchange the information through the assistance of an
AF relay $\Relay$. Each terminal is equipped with a single antenna
which operates in half-duplex mode. We also assume that the channel
for all links induces quasi-static fading, i.e., the channel remains
constant for a block spanning over at least six symbols-intervals
and varies independently for every block. Due to the channel
reciprocity, we note that channel gains of $\mathsf{A}
\longrightarrow \mathsf{B}$ and $\mathsf{B} \longrightarrow
\mathsf{A}$ links, with $\mathsf{A}, \mathsf{B} \in
\{\Source{1},\Source{2},\Relay\}$, are identical.

The communications of the proposed two-way AF DASTC scheme occur in
three phases. In the first phase, both $\Source{1}$ and $\Source{2}$
transmit the first row of Alamouti code \cite{Ala:98:JSAC} to
$\Relay$ with the same transmit power per symbol $\Ps$. The received
signal at the relay $\Relay$ is given by
\begin{align} \label{eq:firstphase}
    \y{R} = \h{1} \s{1} + \h{2} \x{1} + \nR
\end{align}
where $\s{1}= [ {\begin{array}{*{20}c}{s_1 } & {s_2 }  \\
\end{array}}]$ and $\x{1}= [ {\begin{array}{*{20}c}{x_1 } & {x_2 }  \\
\end{array}}]$ are the two first rows of the Alamouti codes generated
by $\Source{1}$ and $\Source{2}$, respectively. The channel
coefficients for the link from $\Source{1}$ and $\Source{2}$ to
$\Relay$, respectively denoted as $\h{1}$ and $\h{2}$, follow a
Rayleigh distribution, i.e., $\h{1} \sim \CG{0}{\CP{1}}$ and $\h{2}
\sim \CG{0}{\CP{2}}$. The vector $\nR$ is the complex additive white
Gaussian noise (AWGN) with zero mean and variance $N_0$.

In the second phase, $\Relay$ amplifies the received signal, i.e.,
$\y{R}$, with the same power constraint as in the first phase and
forwards to $\Source{2}$ while $\Source{1}$ sends the second row of
the Alamouti code to $\Source{2}$. The received signal at the source
$\Source{2}$, $\y{2}$, is written as
\begin{align} \label{eq:secondphase}
   \y{2}
   =
   \h{2} \Gain \y{R} + \h{0} \s{2} + \n{2}
\end{align}
where $\Gain$ is the scalar amplifying gain at the relay, $\h{0}
\sim \CG{0}{\CP{0}}$ is the channel gain for the link between
$\Source{1}$ and $\Source{2}$, $\s{2}= [ {\begin{array}{*{20}c}{-s_2^{\ast} } & {s_1^{\ast} }  \\
\end{array}}]$ is the second row of the Alamouti code, and $\n{2}$ is
the AWGN vector with zero mean and variance $N_0$.

Similarly, in the third phase, $\Relay$ transmit $\Gain \y{R}$
whereas $\Source{2}$ sends the second row of the Alamouti code to
$\Source{1}$. The received signal at the source $\Source{1}$,
$\y{1}$, is shown as
\begin{align} \label{eq:thirdphase}
   \y{1}
   =
   \h{1} \Gain \y{R} + \h{0} \x{2} + \n{1}
\end{align}
where $\x{2}= [ {\begin{array}{*{20}c}{-x_2^{\ast} } & {x_1^{\ast} }  \\
\end{array}}]$ is the second row of Alamouti code and $\n{1}$ is
the AWGN vector with zero mean and variance $N_0$.

The amplifying gain $\Gain$ is determined to satisfy the average
power constraint between the relay and source, i.e., $\Relay$
consumes the same amount of power as each source. In this paper, we
assume channel state information (CSI)-assisted AF relay, i.e., the
relay perfectly knows instantaneous values of the channel gains
$\h{1}$ and $\h{2}$. Taking these information into account, we have
\begin{align}
    \EX{\FN{\Gain \y{R}}^2} = \EX{\FN{\s{i}}^2} =\EX{\FN{\x{i}}^2}
\end{align}
yielding
\begin{align}    \label{eq:Gain:1}
    \Gain^2 = \left(|\h{1}|^2+ |\h{2}|^2 + \frac{1}{\snrave}\right)^{-1}
\end{align}
where $\snrave=\Ps/N_0$ is the average signal-to-noise ratio (SNR).
With sufficient large SNR, the amplifying gain can be tightly
approximated as
\begin{align}    \label{eq:Gain:2}
    \Gain^2 \approx \left(|\h{1}|^2+ |\h{2}|^2 \right)^{-1}
\end{align}
It is assumed that the two sources have perfect knowledge of the
corresponding channel coefficients to fully cancel
self-interference. Hence, the received signal at $\Source{1}$ and
$\Source{2}$ given in \eqref{eq:thirdphase} and
\eqref{eq:secondphase} can readily be formed, respectively, as
follows:
\begin{align} \label{eq:si}
    \y{1}
    &=
    \hnew \B{X} + \z{1}
    \\
    \y{2}
    &=
    \hnew \B{S} + \z{2}
\end{align}
where $\hnew= [ {\begin{array}{*{20}c}{\Gain\h{1}\h{2} } & {\h{0} }  \\
\end{array}}]$, $\z{1}=\Gain \h{1} \nR + \n{1}$, $\z{2}=\Gain \h{2} \nR + \n{2}$, $ \B{S} = \left[ {\begin{array}{*{20}c} {s_1 } & {s_2 }  \\ { - s_2^ *  } & {s_1^ *  }
 \\ \end{array} } \right]$, and $ \B{X} = \left[ {\begin{array}{*{20}c} {x_1 } & {x_2 }  \\ { - x_2^ *  } & {x_1^ *  }
 \\ \end{array} } \right]$. Here, the new noise vector $\z{i}$, with $i \in \{1,2\}$, is the complex Gaussian vector, i.e., $\z{i} \sim \CGV{2}{\B{0}}{N_0\bigl(1+\left|\h{i}\right|^2\Gain^2\bigr)\B{I}_2}$ .
It is important to note that $\B{S}$ and $\B{X}$ are Alamouti
space-time codes constructed at $\Source{1}$ and $\Source{2}$,
respectively.

\section{Average Sum-Rate of Distributed Alamouti Space-Time Codes with Two-Way AF Relaying} \label{sec:SumRate}

In this section, we investigate the information-theoretic
performance of the proposed DASTC with two-way AF relaying presented
in Section~\ref{sec:SystemModel}. As mentioned above, since the
channel model is assumed to be ergodic block fading, we can describe
the average sum-rate as follows:
\begin{align} \label{eq:SumRate:1}
    \Rsum
    &=
    \mathbb{E}
    \biggl\{
    \frac{2}{3}\log_2\det\left(\B{I}_2 + \frac{1}{N_0\left(1+\Gain^2 |\h{2}|^2\right)} \hnew^{\dag}\R{\B{s}}\hnew \right)
    \nonumber
    \\
    &+
    \frac{2}{3}\log_2\det\left(\B{I}_2 + \frac{1}{N_0\left(1+ \Gain^2 |\h{1}|^2\right)} \hnew^{\dag}\R{\B{x}}\hnew \right)
    \biggr\}
\end{align}
where $\R{\B{s}}$ and $\R{\B{x}}$ are the covariance matrices of
Gaussian codewords and the pre-factor $2/3$ accounts for the fact
that each source received information consisting of two symbols over
three time slots. Due to the orthogonality of Alamouti codeword, it
is easy to see that $\R{\B{s}} = \EX{\B{S} \B{S}^{\dag}} = 2\Ps
\B{I}_2$ and $\R{\B{x}} = \EX{\B{X} \B{X}^{\dag}} = 2\Ps \B{I}_2$.
Then we can rewrite \eqref{eq:SumRate:1} as
\begin{align} \label{eq:SumRate:2}
    &\Rsum
    =
    \nonumber
    \\
    &
    \mathbb{E}
    \biggl\{
    \frac{2}{3}\log_2
    \biggl[
    \left(1+2\snrave\frac{|\h{1}|^2 |\h{2}|^2+|\h{0}|^2\left(|\h{1}|^2 + |\h{2}|^2\right)}{|\h{1}|^2 + 2|\h{2}|^2}\right)
    \nonumber
    \\
    &\times
    \left(1+2\snrave\frac{|\h{1}|^2 |\h{2}|^2+|\h{0}|^2\left(|\h{1}|^2 + |\h{2}|^2\right)}{|\h{2}|^2 + 2|\h{1}|^2}\right)
    \biggr]
    \biggr\}
\end{align}
where \eqref{eq:SumRate:2} is established by the fact that
$\det\left(\B{I} + \B{A} \B{B}\right)=\det\left(\B{I} + \B{B}
\B{A}\right)$. For notational simplicity, let us denote
$\snri{i}=2\snrave|\h{i}|^2$ with $i=0,1,2$. Moreover, in the high
SNR regime, i.e., $\snrave$ is sufficient large, the average
sum-rate $\Rsum$ given in \eqref{eq:SumRate:2}, after some
manipulations, can be tightly approximated as
\begin{align} \label{eq:SumRate:3}
    \frac{3 \ln2}{2}
    \Rsum
    &\approx
    \underbrace
    {
    \EX{\ln\left(\frac{\snri{1}\snri{2}}{\snri{1}+2\snri{2}}\right)}
    }_{I_1}
    +
    \underbrace
    {
    \EX{\ln\left(\frac{\snri{1}\snri{2}}{\snri{2}+2\snri{1}}\right)}
    }_{I_2}
    \nonumber
    \\
    &+
    2
    \underbrace
    {
    \EX{\ln\left(1+\frac{\snri{0}\left(\snri{1}+\snri{2}\right)}{\snri{1}\snri{2}}\right)}
    }_{I_3}
\end{align}
In order to calculate \eqref{eq:SumRate:3}, we first start our
derivation by considering the similarity between $I_1$ and $I_2$ in
\eqref{eq:SumRate:3}. To evaluate the integral $I_1$ we utilize the
probability density function (PDF) of $Z$ derived in
Lemma~\ref{lemma:1} given in the Appendix. From \eqref{eq:PDF:Z:1},
we have
\begin{align} \label{eq:I:1:1}
    I_1
    =
    \sum\limits_{u=1}^{2}
    \sum\limits_{v=0}^{1}
    \int\limits_{0}^{\infty}
    \mathcal{A}_v
    \frac{\ln z}{1+z}  z^{u} \exp(-\beta_1 z)
    \Bessel{v}{ \alpha z}
    dz
\end{align}
where
\begin{align*}
    \alpha
    &=
    2\sqrt{\frac{2}{\snravei{1}\snravei{2}}},
    \hspace{0.3cm}
    \beta_1 = \frac{1}{\snravei{2}} + \frac{2}{\snravei{1}}
    \nonumber
    \\
    \mathcal{A}_0 &= \alpha^2,
    \hspace{1.1cm}
    \mathcal{A}_1 = \alpha \beta_1
\end{align*}
To further simplify \eqref{eq:I:1:1}, we will express $\ln z/(1+z)$
and $\exp\left(-\alpha z\right) \Bessel{v}{ \alpha z}$ in terms of
the Meijer's G-function with the help of
\cite[eq.~(8.4.6.13)]{Prudnikov:90:Book} and
\cite[eq.~(6.4.23.3)]{Prudnikov:90:Book} as follows:
\begin{align} \label{eq:ln:1}
    \frac{\ln z}{1 +z}
    &=
    -\pi
    G_{3,3}^{2,2}\left( {z} \left|{\begin{array}{*{20}c} {0,0,1/2}  \vspace{0.2cm}\\
    {0,0,1/2} \end{array}} \right.\right)
    \\ \label{eq:exp:1}
    \exp\left(-\alpha z\right) \Bessel{v}{ \alpha z}
    &=
    \sqrt{\pi}
    G_{1,2}^{2,0}\left({2\alpha z} \left|{\begin{array}{*{20}c} {1/2}  \vspace{0.2cm}\\
    {v,-v} \end{array}} \right. \right)
\end{align}
where $G_{p,q}^{m,n}(\cdot)$ is the Meijer's G-function
\cite[eq.~(8.2.1.1)]{Prudnikov:90:Book}. Moreover, the Meijer's
G-function is a special case of the Fox's H-function
\cite[eq.~(8.3.2.21)]{Prudnikov:90:Book}
\begin{align} \label{eq:Meijer:Fox:1}
    G_{p,q}^{m,n}\left( {z} \left|{\begin{array}{*{20}c} {(a_p)}  \vspace{0.2cm}\\
    {(b_q)} \end{array}} \right.\right)
    =
    H_{p,q}^{m,n}\left( {z} \left|{\begin{array}{*{20}c} {(a_p,1)}  \vspace{0.2cm}\\
    {(b_q,1)} \end{array}} \right.\right)
\end{align}
By combining \eqref{eq:ln:1}, \eqref{eq:exp:1}, and
\eqref{eq:Meijer:Fox:1} with \eqref{eq:I:1:1}, we obtain
\begin{align} \label{eq:I:1:2}
    I_1
    =
    -
    &\sum\limits_{u=1}^{2}
    \sum\limits_{v=0}^{1}
    \pi^{3/2}
    \mathcal{A}_v
    \int\limits_{0}^{\infty}
    z^u
    \exp\left[-\left(\beta_1 - \alpha\right)z\right]
    \nonumber
    \\
    &\times
    H_{3,3}^{2,2}\left[ {z} \left|{\begin{array}{*{20}c} {(0,1),(0,1),(1/2,1)}  \vspace{0.2cm}\\
    {(0,1),(0,1),(1/2,1)} \end{array}} \right.\right]
    \nonumber
    \\
    &\times
    H_{1,2}^{2,0}\left[{2\alpha z} \left|{\begin{array}{*{20}c} {(1/2,1)}  \vspace{0.2cm}\\
    {(v,1),(-v,1)} \end{array}} \right. \right]
    dz
\end{align}
Then, the integral $I_1$ can be calculated with the help of
\cite[eq.~(2.6.2)]{Mathai:78:Book} as follows:
\begin{align} \label{eq:I:1:3}
    &I_1
    =
    -\pi^{3/2}
    \sum\limits_{u=1}^{2}
    \sum\limits_{v=0}^{1}
    \mathcal{A}_v
    \left(\beta_1 - \alpha\right)^{-u-1}
    \nonumber
    \\
    &
H_{1,[3:1],0,[3:2]}^{1,2,0,2,2}\left[ {\begin{array}{*{20}c}
   \frac{1}{\alpha+\beta_1}  \\
   \\
   \\
   \frac{2\alpha}{\alpha+\beta_1}  \\
\end{array}\left| {\begin{array}{*{20}c}
   (1+u,1)  \\
   (1/1,1);(0,1),(0,1),(1/2,1)  \\
   {\_}{\_} \\
   (0,1),(0,1),(\tfrac{1}{2},1);(v,1),(-v,1)  \\
\end{array}} \right.} \right]
\end{align}
where $H_{E,[A:C],F,[B:D]}^{K,N,N',M,M'}[\cdot]$ is the generalized
Fox's $H$-function \cite[eq.~(2.2.1)]{Mathai:78:Book}. Similarly, we
can obtain the closed-form expression for $I_2$.

We next evaluate the integral $I_3$ by utilizing the PDF of $T$
derived in Lemma~\ref{lemma:2} given in the Appendix. From
\eqref{eq:PDF:T:1}, we have
\begin{align} \label{eq:I:3:1}
    I_3
    =
    \sum\limits_{n=1}^2
    \xi_n
    &\int\limits_{0}^{\infty}
    \ln\left(1+t\right)
    \left(t+\zeta\right)^{-3-n}
    \nonumber
    \\
    &\times
    {_2}F_1\left(3+n,n+\tfrac{1}{2};\tfrac{7}{2};\tfrac{t+\eta}{t+\zeta}\right)
    dt
\end{align}
Then we exchange the variable $\frac{t+\eta}{t+\zeta}=\frac{x}{x+1}$
and assume that the relaying protocol is symmetric, i.e.,
$\snravei{1}=\snravei{2}$, yielding $\eta=0$. We next express
$\ln(1+ \zeta x)$ and $\left(1+x\right)^{-a}
{_2}F_1\left(a,b;c;\tfrac{x}{x+1}\right)$ in terms of the Meijer's
G-function with the help of \cite[eq.~(8.4.6.5)]{Prudnikov:90:Book}
and \cite[eq.~(8.4.49.26)]{Prudnikov:90:Book} as follows:
\begin{align} \label{eq:ln:2}
    \ln(1+\zeta x)
    &=
    G_{2,2}^{1,2}\left( {\zeta x} \left|{\begin{array}{*{20}c} {1,1}  \vspace{0.2cm}\\
    {1,0} \end{array}} \right.\right)
    \\ \label{eq:GaussHyper:1}
    \left(1+x\right)^{-3-n}
    &{_2}F_1\left(3+n,n+\tfrac{1}{2};\tfrac{7}{2};\tfrac{x}{x+1}\right)
    \nonumber
    \\
    &=
    G_{2,2}^{1,2}\left( {x} \left|{\begin{array}{*{20}c} {-2-n,\tfrac{1}{2}-n}  \vspace{0.2cm}\\
    {0,-2n} \end{array}} \right.\right)
\end{align}
By combining \eqref{eq:ln:2} and \eqref{eq:GaussHyper:1} with
\eqref{eq:I:3:1}, we have
\begin{align} \label{eq:I:3:2}
    I_3
    =
    \sum\limits_{n=1}^2
    \xi_n
    \int\limits_{0}^{\infty}
    &G_{2,2}^{1,2}\left( {x} \left|{\begin{array}{*{20}c} {-2-n,\tfrac{1}{2}-n}  \vspace{0.2cm}\\
    {0,-2n} \end{array}} \right.\right)
    \nonumber
    \\
    &\times
    G_{2,2}^{1,2}\left( {\zeta x} \left|{\begin{array}{*{20}c} {1,1}  \vspace{0.2cm}\\
    {1,0} \end{array}} \right.\right)
    dx
\end{align}
The integral given in \eqref{eq:I:3:2} can be finalized in the form
of Fox's H-function by using \cite[eq.~(7.811.1)]{GR:00:Book} and
\cite[eq.~(8.3.2.21)]{Prudnikov:90:Book} yielding $I_3$ as follows:
\begin{align} \label{eq:I:3:3}
    I_3
    =
    \sum\limits_{n=1}^2
    \xi_n
    H_{4,4}^{4,2}\left[{\zeta} \left|{\begin{array}{*{20}c} {(0,1),(2n,1),(1,1),(1,1)}  \vspace{0.2cm}\\
    {(2+n,1),(n-\tfrac{1}{2},1),(1,1),(0,1)} \end{array}} \right.
    \right]
\end{align}
By substituting \eqref{eq:I:1:3} and $\eqref{eq:I:3:3}$ in
\eqref{eq:SumRate:3} and considering the fact that $I_2$ can be
obtained in a similar form of $I_1$, the sum-rate of DASTC with
two-way AF relay can be shown as
\begin{align}
    &\Rsum
    \approx
    -\frac{2\pi^{3/2}}{3\ln 2}
    \sum\limits_{k=1}^2
    \sum\limits_{u=1}^{2}
    \sum\limits_{v=0}^{1}
    \mathcal{A}_v
    \left(\beta_k - \alpha\right)^{-u-1}
    \nonumber
    \\
    &
H_{1,[3:1],0,[3:2]}^{1,2,0,2,2}\left[ {\begin{array}{*{20}c}
   \frac{1}{\alpha+\beta_k}  \\
   \\
   \\
   \frac{2\alpha}{\alpha+\beta_k}  \\
\end{array}\left| {\begin{array}{*{20}c}
   (1+u,1)  \\
   (1/1,1);(0,1),(0,1),(1/2,1)  \\
   {\_}{\_} \\
   (0,1),(0,1),(\tfrac{1}{2},1);(v,1),(-v,1)  \\
\end{array}} \right.} \right]
    \nonumber
    \\
    &+
    \frac{4}{3\ln2}
    \!
    \sum\limits_{n=1}^2
    \xi_n
    H_{4,4}^{4,2} \! \left[{\zeta} \left|{\begin{array}{*{20}c} {(0,1),(2n,1),(1,1),(1,1)}  \vspace{0.2cm}\\
    {(2+n,1),(n-\tfrac{1}{2},1),(1,1),(0,1)} \end{array}} \right.
    \right]
\end{align}
\section{Numerical Results} \label{sec:Result}

In this section, we provide the numerical results to verify the
proposed two-way DASTC and the correctness of our analysis in two
specific examples. The path loss of each link follows an
exponential-decay model: if the distance between the two sources is
equal to $d$, then $\CP{0} \propto d^{-\alpha}$ where an exponent of
$\alpha=4$ corresponds to a typical non line-of-sight propagation
\cite{HTHC:08:VTCSpring}. Here we assume that the
$\Source{1}\longrightarrow \Source{2}$ link has unit channel mean
power, i.e., $\CP{0}=1$. In the first example, we consider the
symmetric relaying protocol, i.e., $\CP{0}=\CP{1}=\CP{2}=1$. In the
second example, we assume that the geometry for locations of three
communicating terminals is co-linear where the relay is placed half
way between the two sources, i.e., $\CP{1}=\CP{2}=16$ and
$\CP{0}=1$.

\begin{figure}[t]
    \centerline{\includegraphics[width=0.42\textwidth]{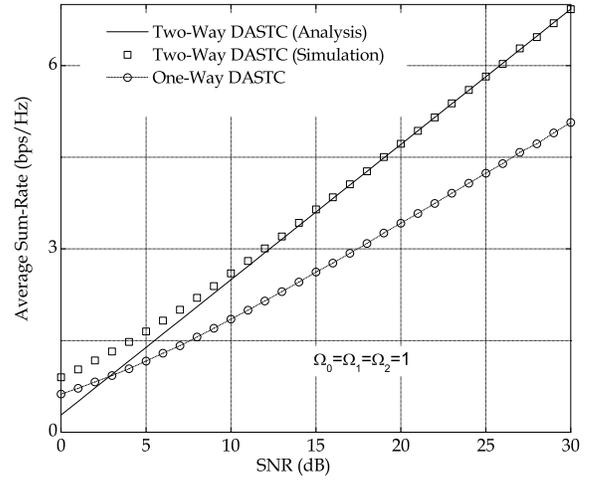}}
    \caption{
        Average sum-rate of DASTC with symmetric relaying protocol.
    }
    \label{fig:1}
\end{figure}

\begin{figure}[t]
    \centerline{\includegraphics[width=0.42\textwidth]{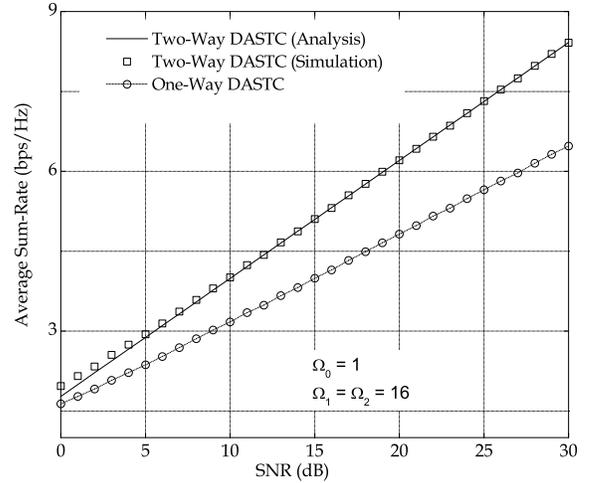}}
    \caption{
        Average sum-rate of DASTC with co-linear relaying protocol.
    }
    \label{fig:2}
\end{figure}

Fig.~\ref{fig:1} and Fig.~\ref{fig:2} illustrate the average
sum-rate of DASTC in two-way AF relay networks versus average SNR
for the two considered examples. As can be observed from these two
figures, the analysis is very tight from the middle to high SNR
regime. Specifically, from SNR=10 dB the analytical and simulation
curves perfectly match with each other which verify the tightness of
our approximation.

More importantly, the proposed two-way DASTC scheme outperforms the
conventional one-way system in terms of the spectral efficiency in
the whole considered range of SNR. In particular, at SNR=30 dB, the
proposed scheme enhances the average sum-rate to 3.8 bps/Hz in both
Example 1 and 2 compared to the conventional one. It is interesting
to observe that with a fixed value of $\CP{0}$ the gain is
unchanged, irrespective of the relay's location.
\section{Conclusions} \label{sec:Conclusions}

We have proposed a DASTC scheme for two-way AF relay networks that
circumvents the loss in spectral efficiency inherently occurred in
conventional one-way DASTC system. We also derive the tight
approximation for the average sum-rate of the proposed scheme. The
closed-form expression for approximated sum-rate is given in the
form of Fox's H-function which readily allows us to analyze the
spectral efficiency of the proposed scheme. The numerical results
provided have validated our analysis.


\appendix
\section{Auxiliary Results} \label{appendix:1}

The following lemmas will be helpful in the paper.

\begin{lemma} \label{lemma:1}

Let $\snri{i}$ and $\snri{j}$ be statistically independent and not
necessarily identically distributed (i.n.i.d.) exponential random
variables with hazard rate $\snravei{i}$ and $\snravei{j}$,
respectively. Suppose that the ratio of $Z$ takes the form
\begin{align}
    Z=\frac{\snri{i} \snri{j}}{\snri{i} +2\snri{j}}
\end{align}
Then, we obtain the PDF of random variable $Z$ as
\begin{align} \label{eq:PDF:Z:1}
    \PDF{Z}{z}
    =
    \alpha z \exp\left(-\beta_i z\right)
    \left[\alpha\Bessel{0}{\alpha z} + \beta_i \Bessel{1}{\alpha z}\right]
\end{align}
where $\alpha = 2\sqrt{\frac{2}{\snravei{i}\snravei{j}}}$ and
$\beta_i = \frac{1}{\snravei{j}} + \frac{2}{\snravei{i}}$.
\begin{proof}
Let us rewrite $Z$ in a more tractable form as
\begin{align} \label{eq:Z}
    Z
    =
    \left(X + Y\right)^{-1}
\end{align}
where $X=1/\snri{i}$ and $Y=2/\snri{j}$. Then, after some algebraic
simplifications together with the help of
\cite[eq.~(3.471.9)]{GR:00:Book}, the moment generating function
(MGF) of $X$ and $Y$ can be expressed as, respectively
\begin{align} \label{eq:MGF:X}
    &\MGF{X}{s}
    =
    \EXs{X}{\exp\left(-s x\right)}
    =
    2 \sqrt{\frac{s}{\snravei{i}}} \Bessel{1}{2
    \sqrt{\frac{s}{\snravei{i}}}}
    \\ \label{eq:MGF:Y}
    &\MGF{Y}{s}
    =
    \EXs{Y}{\exp\left(-s y\right)}
    =
    2 \sqrt{\frac{2s}{\snravei{j}}} \Bessel{1}{2
    \sqrt{\frac{2s}{\snravei{j}}}}
\end{align}
As can be seen from \eqref{eq:Z}, since $1/Z$ is the sum of two
statistically independent random variables, by using
\eqref{eq:MGF:X} and \eqref{eq:MGF:Y}, the MGF of $1/Z$ can be
expressed as
\begin{align} \label{eq:MGF:1:Z}
    \MGF{1/Z}{s}
    =
    4\sqrt{\frac{2}{\snravei{i}\snravei{j}}}s \Bessel{1}{2
    \sqrt{\frac{s}{\snravei{i}}}}
    \Bessel{1}{2
    \sqrt{\frac{2s}{\snravei{j}}}}
\end{align}
Then, the cumulative distribution function (CDF) of $1/Z$,
$\CDF{1/Z}{x}$, can be shown as
\begin{eqnarray} \label{eq:CDF:1:Z}
    \CDF{1/Z}{x}
     =1- \mathcal{L}^{-1}\{\MGF{1/Z}{s}/s\}|_{1/x}
\end{eqnarray}
where $\mathcal{L}^{-1}\{.\}$ stands for the inverse Laplace
transform. From \eqref{eq:MGF:1:Z} and \eqref{eq:CDF:1:Z} together
with \cite[eq.~(13.2.20)]{RK:66:Book}, we have
%
%
\begin{align} \label{eq:CDF:Z:1}
    \CDF{Z}{z}
    =
    1 - \alpha z \exp\left(-\beta_i z\right)
    \Bessel{1}{\alpha z}
\end{align}
where $\alpha = 2\sqrt{\frac{2}{\snravei{i}\snravei{j}}}$ and
$\beta_i = \frac{1}{\snravei{j}} + \frac{2}{\snravei{i}}$. By
differentiating \eqref{eq:CDF:Z:1} with respect to $z$ and using the
fact that
$\tfrac{d\Bessel{v}{z}}{dz}=-\Bessel{v-1}{z}-\tfrac{v}{z}\Bessel{v}{z}$
\cite[eq.~(8.486.12)]{GR:00:Book}, we obtain \eqref{eq:PDF:Z:1}
which completes the proof.
\end{proof}
\end{lemma}

\begin{lemma} \label{lemma:2}
Let $\snri{0}$, $\snri{1}$, and $\snri{2}$ be the three i.n.i.d.
exponential random variables with hazard rates $\snravei{0}$,
$\snravei{1}$, and $\snravei{2}$, respectively, then the PDF of
$T=\snri{0}\frac{\snri{1}+\snri{2}}{\snri{1}\snri{2}}$ is given by
\begin{align} \label{eq:PDF:T:1}
    \PDF{T}{t}
    =
    \! \!
    \sum\limits_{n=1}^2
    \xi_n
    \tfrac{1}{(t+\zeta)^{3+n}}
    {_2}F_1\left(3+n,n+\tfrac{1}{2};\tfrac{7}{2};\tfrac{t+\eta}{t+\zeta}\right)
\end{align}
where
\begin{align*}
    \xi_n 
    &=
    \sqrt{\pi}
    \left(\frac{4}{\sqrt{\snravei{1}\snravei{2}}}\right)^{n}
    \frac{\Gamma\left(3+n\right)\Gamma\left(3-n\right)}{\Gamma\left(7/2\right)}
    \varpi_n
    \nonumber
    \\
    \varpi_0 
    &=
    \frac{4}{\snravei{0}\snravei{1}\snravei{2}},
    \hspace{0.5cm}
    \varpi_1
    =
    \frac{2\left(\snravei{1}+\snravei{2}\right)}{\snravei{0}\left(\snravei{1}\snravei{2}\right)^{3/2}}
    \nonumber
    \\
    \zeta 
    &=
    \snravei{0} \left(\frac{1}{\snravei{1}} + \frac{1}{\snravei{2}} +
    \frac{2}{\snravei{1}\snravei{2}}\right),
    \hspace{0.3cm}
    \eta 
    =
    \snravei{0} \left(\frac{1}{\snravei{1}} + \frac{1}{\snravei{2}} - \frac{2}{\snravei{1}\snravei{2}}\right)
\end{align*}
and $_2F_1\left(a,b;c;z\right)$ is the Gauss hypergeometric function
\cite[eq.~(2.12.1)]{Erd:53:Book}.
\begin{proof} Since $T=\snri{0}\frac{\snri{1}+\snri{2}}{\snri{1}\snri{2}}$, by defining $W=\frac{\snri{1}\snri{2}}{\snri{1}+\snri{2}}$, we have
\begin{align} \label{eq:CDF:T:1}
    \CDF{T}{t}
    =
    \int\nolimits_{0}^{\infty}
    \left[1-\exp\left(-\frac{wt}{\snravei{0}}\right)\right]
    \PDF{W}{w} dw
\end{align}
Following similar steps as in Lemma~\ref{lemma:1}, we obtain
$\PDF{W}{w}$ as
\begin{align} \label{eq:PDF:W:1}
    &\PDF{W}{w}
    =
    \frac{2w}{\snravei{1}\snravei{2}}
    \exp\left[-\left(\frac{1}{\snravei{1}}+\frac{1}{\snravei{2}}\right)w\right]
    \nonumber
    \\
    &\times
    \left[
    2\Bessel{0}{\frac{2w}{\sqrt{\snravei{1}\snravei{2}}}}
    +
    \left(\frac{\snravei{1}+\snravei{2}}{\snravei{1}\snravei{2}}\right)
    \Bessel{1}{\frac{2w}{\sqrt{\snravei{1}\snravei{2}}}}
    \right]
\end{align}
By substituting \eqref{eq:PDF:W:1} in \eqref{eq:CDF:T:1} and taking
the derivative with respect to $t$, we obtain the PDF of $T$ as
\begin{align} \label{eq:PDF:T:2}
    \PDF{T}{t}
    =
    \sum\limits_{n=1}^{2}
    \varpi_n
    \int\limits_{0}^{\infty}
    w^2
    &\exp\left[-\left(\frac{t}{\snravei{0}}+\frac{1}{\snravei{1}}+\frac{1}{\snravei{2}}\right)w\right]
    \nonumber
    \\
    &\times
    \Bessel{n}{\frac{2w}{\sqrt{\snravei{1}\snravei{2}}}}
    dw
\end{align}
It is observed that the integral given in \eqref{eq:PDF:T:2} can be
simplified by using \cite[eq.~(6.621.3)]{Prudnikov:90:Book} which
results in \eqref{eq:PDF:T:1}. This completes the proof.
\end{proof}
\end{lemma}

\bibliographystyle{IEEEtran}
\bibliography{IEEEabrv,TwoWayAlamouti}
\end{document}